\documentclass[conference,compsoc]{IEEEtran}
\IEEEoverridecommandlockouts
\usepackage{cite}
\usepackage{subcaption}
\usepackage[ruled,vlined,linesnumbered]{algorithm2e}
\usepackage{amsmath,amssymb,amsfonts}
\usepackage{algorithmic}
\usepackage{graphicx}
\usepackage{url}
\usepackage[hidelinks]{hyperref}
\usepackage{textcomp}
\usepackage{tcolorbox}
\usepackage{xcolor}
\usepackage[table, svgnames]{xcolor} 
\def\BibTeX{{\rm B\kern-.05em{\sc i\kern-.025em b}\kern-.08em
    T\kern-.1667em\lower.7ex\hbox{E}\kern-.125emX}}

\usepackage{booktabs}
\usepackage[nonumberlist,nogroupskip,nopostdot]{glossaries}

\begin{document}
\newacronym{vnf}{VNF}{Virtualized Network Function}
\newacronym{5g}{5G}{Fifth Generation}
\newacronym{5g-advanced}{5G-Advanced}{5G-Advanced}
\newacronym{6g}{6G}{Sixth Generation}
\newacronym{mmtc}{mMTC}{massive Machine Type Communications}
\newacronym{embb}{eMBB}{enhanced Mobile Broadband}
\newacronym{urllc}{URLLC}{Ultra-Reliable Low Latency Communications}
\newacronym{sdn}{SDN}{Software Defined Networking}
\newacronym{nfv}{NFV}{Network Function Virtualization}
\newacronym{nf}{NF}{Network Function}
\newacronym{dn}{DN}{Data Network}
\newacronym{arpf}{ARPF}{Authentication credential Repository and Processing Function}
\newacronym{gtp}{GTP}{General Packet Radio Service Tunnelling Protocol}
\newacronym{gprs}{GPRS}{General Packet Radio Service}
\newacronym{teid}{TEID}{Tunnel Endpoint Identifier}
\newacronym{seid}{SEID}{Session Endpoint Identifier}

\newacronym{ip}{IP}{Internet Protocol}
\newacronym{icmp}{ICMP}{Internet Control Message Protocol}
\newacronym{dns}{DNS}{Domain Name System}
\newacronym{pcap}{PCAP}{Packet Capture}
\newacronym{pps}{PPS}{Packets Per Second}
\newacronym{rto}{RTO}{Retransmission Timeout}
\newacronym{cwnd}{CWND}{Congestion Window}
\newacronym{sba}{SBA}{Service-Based Architecture}

\newacronym{ietf}{IETF}{Internet Engineering Task Force}

\newacronym{supi}{SUPI}{Subscription Permanent Identifier}
\newacronym{capex}{CAPEX}{Capital Expenditure}
\newacronym{opex}{OPEX}{Operational Expenditure}
\newacronym{dos}{DoS}{Denial-of-Service}
\newacronym{ddos}{DDoS}{Distributed DoS}
\newacronym{minlp}{MINLP}{Mixed-Integer Nonlinear Programming}
\newacronym{milp}{MILP}{Mixed-Integer Linear Programming}
\newacronym{ilp}{ILP}{Integer Linear Programming}
\newacronym{upf}{UPF}{User Plane Function}
\newacronym{mec}{MEC}{Multi-access Edge Computing}
\newacronym{qos}{QoS}{Quality of Service}
\newacronym{qoe}{QoE}{Quality of Experience}
\newacronym{sfc}{SFC}{Service Function Chain}
\newacronym{pdu}{PDU}{Protocol Data Unit}
\newacronym{ns}{NS}{Network Slicing}
\newacronym{sla}{SLA}{Service Level Agreement}
\newacronym{slo}{SLO}{Service Level Objective}
\newacronym{dps}{DPS}{Data Plane Services}
\newacronym{cps}{CPS}{Control Plane Services}
\newacronym{dp}{DP}{Data Plane}
\newacronym{cp}{CP}{Control Plane}
\newacronym{vm}{VM}{Virtual Machine}
\newacronym{mip}{MIP}{Mixed Integer Programming}
\newacronym{amf}{AMF}{Access and Mobility Function}
\newacronym{smf}{SMF}{Session Management Function}
\newacronym{nrf}{NRF}{Network Repository Function}
\newacronym{scip}{SCIP}{Solving Constraint Integer Programs}
\newacronym{ue}{UE}{User Equipment}

\newacronym{hplmn}{H-PLMN}{Home \gls{plmn}}
\newacronym{plmn}{PLMN}{Public Land Mobile Network}
\newacronym{eap-aka}{EAP-AKA'}{Extensible Authentication Protocol-Authentication and Key Agreement}
\newacronym{aka}{AKA}{Authentication and Key Agreement}
\newacronym{gnb}{gNB}{gNodeB}
\newacronym{3gpp}{3GPP}{3rd Generation Partnership Project}
\newacronym{cn}{CN}{Core Network}
\newacronym{ran}{RAN}{Radio Access Network}
\newacronym{udm}{UDM}{Unified Data Management}
\newacronym{ausf}{AUSF}{Authentication Server Function}

\newacronym{mno}{MNO}{Mobile Network Operator}

\newacronym{udp}{UDP}{User Datagram Protocol}
\newacronym{tcp}{TCP}{Transmission Control Protocol}

\newacronym{rtt}{RTT}{Round-Trip Time}

\newacronym{v2x}{V2X}{Vehicle-to-Everything}
\newacronym{dpdk}{DPDK}{Data Plane Development Kit}
\newacronym{gcn}{GCN}{Graph Convolutional Network}
\newacronym{nsga}{NSGA-II}{Non-Dominated Sorting Genetic Algorithm II}

\newacronym{iot}{IoT}{Internet of Things}
\newacronym{b2b2x}{B2B2X}{Business-to-Business-to-Everything}
\newacronym{vpn}{VPN}{Virtual Private Network}

\newacronym{ccs}{CCS}{Conference on Computer and Communications Security}
\newacronym{vlan}{VLAN}{Virtual Local Area Network}
\newacronym{pfcp}{PFCP}{Packet Forwarding Control Protocol}
\newacronym{uav}{UAV}{Unmanned Aerial Vehicle}
\newacronym{xr}{XR}{Extended Reality}

\newacronym{xdp-ddm}{XDP-DDM}{XDP Detection \& Drop Module}
\newacronym{up-ta}{UP-TA}{User-Plane Telemetry Agent}
\newacronym{cp-mo}{CP-MO}{Control-Plane Mitigation Orchestrator}
\newacronym{ebpf}{eBPF}{extended Berkeley Packet Filter}
\newacronym{xdp}{XDP}{eXpress Data Path}
\newacronym{api}{API}{Application Programming Interface}
\newacronym{udr}{UDR}{Unified Data Repository}

\title{Mitigating Nested GTP Attacks: A Closed-Loop eBPF Defense for 5G and Beyond Networks\\
{\large Author’s draft for soliciting feedback -- \today}
}

\author{\IEEEauthorblockN{AbdulAziz AbdulGhaffar}
\IEEEauthorblockA{\textit{Department of Systems and Computer Engineering} \\
\textit{Carleton University}\\
Ottawa, Canada 
}
\and
\IEEEauthorblockN{Ashraf Matrawy}
\IEEEauthorblockA{\textit{School of Information Technology} \\
\textit{Carleton University}\\
Ottawa, Canada 
}
}

\maketitle

\begin{abstract}
The evolving cellular networks continue to rely on the legacy GPRS Tunneling Protocol-User Plane (GTP-U) to encapsulate and route user data traffic across the data plane. Recent security research demonstrated that nested GTP-U (GTP-in-GTP) payload vulnerabilities enable a diverse range of attacks on victim users. To address this vulnerability, in this paper, we propose a closed-loop extended Berkeley Packet Filter (eBPF) and eXpress Data Path (XDP) defense to detect and mitigate nested GTP-U attacks in 5G and beyond networks. The proposed framework contains three components: an in-kernel XDP Detection and Drop Module (XDP-DDM), a user-space User-Plane Telemetry Agent (UP-TA), and a multi-threaded Control-Plane Mitigation Orchestrator (CP-MO).

We implement and evaluate the proposed defense on a sliced 5G testbed integrating Free5GC, UERANSIM, and Containerlab. Experimental results indicate that the XDP-DDM detects and drops malicious nested GTP-U packets at line rate, while the UP-TA and CP-MO asynchronously report and revoke the attacker's network access, respectively. Under a 10,000 packets-per-second cross-slice nested GTP-U flood, the proposed solution preserves the baseline performance of the victim slices while introducing minimal overhead. We also show that the proposed solution has minimal impact on the user’s throughput and round-trip time.
\end{abstract}

\begin{IEEEkeywords}
5G and 6G security, closed-loop mitigation, eBPF, XDP, GTP-U encapsulation, network slicing.
\end{IEEEkeywords}

\section{Introduction} \label{sec:intro}
As cellular networks evolve from \gls{5g} towards \gls{6g} networks, recent \gls{3gpp} Release 20 standardization studies on the \gls{6g} system architecture, \gls{3gpp} TR 23.801-1~\cite{3gpp.23.801}, establish that \gls{6g} will adopt the current \gls{5g} \gls{sba}, Control and User Plane Separation (CUPS), and network slicing as its baseline starting point~\cite{3gpp.23.801}. However, the \gls{dp} of cellular networks relies on the legacy \gls{gtp} User Plane (GTP-U)~\cite{3gpp.29.281}.

Recently, Shaik~\textit{et al.}~\cite{shaik2025uncovering, shaik2025titanic} demonstrated at ACM \gls{ccs} 2025 that a compromised \gls{ue}, operating with only a smartphone and a valid SIM, can exploit the vulnerabilities in \gls{gtp}-U packet processing at the \gls{upf} to execute severe protocol tunneling and network boundary bridging attacks across six \gls{5g} cores, including commercial networks. Because emerging \gls{6g} architectures are expected to preserve \gls{gtp}-U tunneling for user-plane forwarding across the N3 and N9 interfaces~\cite{3gpp.23.801}, these underlying protocol vulnerabilities represent a persistent threat that will be directly inherited by \gls{6g} networks unless automated \gls{dp} defenses are established. 

\subsection{Problem Statement}
\gls{gtp}-U was originally standardized in 2.5G/3G networks and was engineered for high-throughput forwarding rather than adversarial resilience~\cite{shaik2025uncovering, zhang2025invade}. In a standard \gls{5g} deployment, uplink user traffic is encapsulated by the serving \gls{gnb} into an outer \gls{gtp}-U header containing the destination \gls{upf}'s \gls{ip} address and a 32-bit \gls{teid} for the \gls{ue}~\cite{3gpp.29.281, shaik2025uncovering}. Because \gls{gtp}-U lacks source authentication or recursive payload validation, standard \gls{upf} implementations validate only the outermost \gls{gtp}-U header and treat the inner encapsulated payload as an opaque byte stream~\cite{shaik2025titanic, shaik2025uncovering}.

As Shaik~\textit{et al.}~\cite{shaik2025uncovering} demonstrated, an authenticated adversary can abuse this blind decapsulation by injecting a crafted packet containing a nested \gls{gtp}-U header (\gls{gtp}-in-\gls{gtp}) into their legitimate uplink tunnel. When the \gls{upf} strips the legitimate outer \gls{gtp}-U header, it fails to inspect the inner protocol layer and blindly routes the inner forged \gls{gtp}-U packet back into the internal user-plane or control-plane interfaces. While Shaik~\textit{et al.}~\cite{shaik2025uncovering} demonstrated these exploits in unsliced cores, the threat is further magnified in multi-tenant \gls{5g}/\gls{6g} network slicing environments~\cite{aziz2026slice}.

\subsection{Motivation}
Mitigating nested \gls{gtp}-U attacks is critical due to the severity of their impact on legitimate subscribers and the multi-vector exploitation~\cite{TrendAI_5g}. Rather than resulting in a single attack, nested \gls{gtp}-U encapsulation functions as a highly exploitable mechanism for ``Protocol Tunneling'', a critical threat category formally recognized by the MITRE FiGHT framework~\cite{mitre2025fight}. By varying the inner encapsulated payload, an adversary can execute: \textcircled{1} \textit{\gls{teid} and \gls{seid} enumeration}~\cite{shaik2025uncovering}, \textcircled{2} \textit{direct \gls{ue}-to-\gls{ue} packet injection}~\cite{shaik2025uncovering}, \textcircled{3} \textit{control-plane boundary bridging} (\gls{pfcp}-in-GTP)~\cite{kwon2026mobile,shaik2025uncovering}, \textcircled{4} \textit{amplified reflection attacks} causing billing fraud and quota depletion~\cite{shaik2025uncovering,zhang2025invade}, and \textcircled{5} \textit{cross-slice injection}~\cite{aziz2026slice}. Hence, neutralizing nested \gls{gtp}-U encapsulation at the \gls{upf} ingress is crucial for eliminating these attack vectors.

In this work, we focus on neutralizing nested \gls{gtp}-U-in-\gls{gtp}-U encapsulation at the \gls{upf} ingress, while noting that the same \gls{xdp} parsing pipeline can be directly extended to detect control-plane boundary bridging, such as \gls{pfcp}-in-\gls{gtp}.

\subsection{Research Questions}
This work investigates how to construct an autonomous, closed-loop, zero-trust defense directly within the \gls{upf} \gls{dp} and \gls{5g} core \gls{cp}. Specifically, we address three core research questions (RQs) that directly guide our architectural design goals and experimental evaluation:

    \textbf{RQ1 (Data-Plane Enforcement):} \textit{How can nested \gls{gtp}-U encapsulation attacks be deterministically detected and dropped at wire speed within the \gls{upf} without imposing significant performance penalties on legitimate traffic?}
    \begin{itemize}
        \item Addressed by our \textit{`Lightweight Processing'} and \textit{`Plug-and-Play Integration'} design goals via the in-kernel \acrfull{xdp-ddm} in Section~\ref{sec:xdp_ddm}, and validated in Section~\ref{sec_result_impact} and \ref{sec_result_latency}.
    \end{itemize}
    
    \textbf{RQ2 (Non-Blocking Cross-Plane Telemetry):} \textit{How can high-volume attack detections in the kernel be communicated to the user-space and across the N4 interface without inducing alert storms or CPU bottlenecks during a volumetric flood?}
    \begin{itemize}
        \item Addressed by our \textit{`High Reliability'} design goal via a 10-second BPF Least Recently Used (LRU) map debounce mechanism and the \gls{up-ta} in Sections~\ref{sec:xdp_ddm} \& \ref{sec:up_ta}, and quantified in Section~\ref{sec_result_latency}.
    \end{itemize}
    
    \textbf{RQ3 (Automated Closed-Loop Session Revocation):} \textit{How can the \gls{5g}/\gls{6g} \gls{cp} autonomously translate \gls{dp} threat telemetry into synchronized, permanent \gls{ue} session termination and blacklisting across \gls{3gpp} \gls{sba} functions while preserving victim slice resilience under active flooding?}
     \begin{itemize}
        \item Addressed by the multi-threaded \gls{cp-mo} interacting with \gls{5g} \gls{cp} \glspl{nf} in Section~\ref{sec:cp_mo}, and evaluated under a 10,000 \gls{pps} flood in Section~\ref{sec_result_resilience}.
    \end{itemize}

\subsection{Contributions}
The main contributions of this paper are summarized as follows:
\begin{enumerate}
    \item \textbf{Closed-Loop \acrshort{ebpf}/\acrshort{xdp} Security Defense:} We propose a novel Closed-Loop defense framework that enables kernel-space \acrshort{dp} threat mitigation along with automated \acrshort{cp} attacker revocation to defeat nested \acrshort{gtp}-U attacks. \textcircled{a} \textbf{Line-Rate Detection \& Telemetry:} We develop the in-kernel \acrshort{ebpf}/\acrshort{xdp} program (\acrshort{xdp-ddm}) attached to the \acrshort{upf} N3 ingress interface that inspects nested \acrshort{gtp} headers and drops malicious packets in $<0.4\ \mu\text{s}$. Using a BPF LRU map for 10-second per-attacker alert debouncing, a user-space agent \acrshort{up-ta} dispatches lightweight JSON alerts to the \acrshort{cp} within $\sim 1 \text{ ms}$ without bottlenecking the \acrshort{upf} CPU. \textcircled{b} \textbf{Automated CP Revocation:} We engineer a multi-threaded \acrshort{cp-mo} agent which resolves malicious inner IPs to their \acrshort{supi} and executes a \acrshort{3gpp} REST \acrshort{api} teardown across the \acrshort{smf}, \acrshort{amf}, and \acrshort{udr} to permanently blacklist the adversary.
    \item \textbf{Plug-and-Play Compatibility for \acrshort{5g}/\acrshort{6g}:} Our \acrshort{xdp-ddm} solution operates natively within the Linux network driver before packets reach the \acrshort{upf} application; it is completely decoupled from the underlying \acrshort{upf} implementation.  It can be seamlessly deployed as a plug-and-play security layer on any Linux-based \acrshort{upf} across current \acrshort{5g} and future \acrshort{6g} architectures without modifying existing \acrshort{vnf} source code.
    \item \textbf{Empirical Validation on a Sliced 5G Testbed:} We deploy and evaluate the proposed defense on a \acrshort{5g} network slicing testbed integrating Free5GC, UERANSIM, and Containerlab. Under a volumetric cross-slice nested \acrshort{gtp}-U flood, our defense neutralizes the attack and preserves the victim's legitimate traffic from the severe impacts observed with the baseline \acrshort{upf}. To foster reproducibility, we release our core \acrshort{ebpf}/\acrshort{xdp} modules, telemetry/orchestration agents, and deployment scripts as open-source artifacts\footnote{The source code will be made publicly available upon publication at: \url{https://github.com/abdulazizag/Nested-GTP-Mitigation}}.
\end{enumerate}

The remainder of this paper is organized as follows. Section~\ref{sec:related} reviews the related work and establishes the research gap. Section~\ref{sec:threat_model} defines the system assumptions and threat model, and explains the attack, while Section~\ref{sec:proposed} details the proposed closed-loop \gls{ebpf}/\gls{xdp} defense. Section~\ref{sec:methodology} outlines the \gls{5g} testbed environment and evaluation methodology, followed by a comprehensive analysis of the experimental results in Section~\ref{sec:results}. The limitations of our work are discussed in Section~\ref{sec:limitation}, and Section~\ref{sec:ethical} outlines the ethical considerations of our work. Section~\ref{sec:conclusion} concludes the paper.

\section{Related Works} \label{sec:related}
This section reviews the state of the art across three domains relevant to our work. Table~\ref{tab:related_work} summarizes the most closely related studies and highlights how our proposed closed-loop \gls{ebpf}/\gls{xdp} defense advances the state of the art.

\begin{table*}[t!]
\centering
\caption{Comparison of Related Works Across 5G User-Plane Security, Slice Isolation, and Programmable Data Planes}
\label{tab:related_work}
\renewcommand{\arraystretch}{1.15}
\setlength{\tabcolsep}{4.5pt}
\resizebox{\textwidth}{!}{
\begin{tabular}{l l p{3.0cm} c c c c}  \toprule[1.5pt]
\rowcolor{gray!20}
\textbf{Reference} & \textbf{Primary Scope} & \textbf{Defense Paradigm} & \textbf{Nested GTP-U} & \textbf{Line-Rate Drop}  & \textbf{Automated 3GPP} & \textbf{Empirical 5G} \\
\rowcolor{gray!20}
 &  &  & \textbf{Inspection} & \textbf{($\mu\text{s}$-level XDP)} & \textbf{UE Revocation} & \textbf{Attack Testbed} \\
 \toprule[1.5pt]
Salim (Trend Micro) \cite{salim2023attacks} & 5G User Plane (N3) & CT-aware DPI & \checkmark & $\times$ &  $\times$ & $\times$ \\
\rowcolor{blue!7}
Zhang \textit{et al.} \cite{zhang2025invade} & Exposed 4G/5G GTP & Protocol Auth. / Auditing & $\times$ & $\times$ & $\times$ & \checkmark (Open5GS/free5GC) \\
Kwon \textit{et al.} \cite{kwon2026mobile} & 5G SA Core (N3/N4) & Routing / ACL Hardening & \checkmark (PFCP-in-GTP) & $\times$ & $\times$ & \checkmark (Carrier / 5G SA) \\
\rowcolor{blue!7}
Shaik \textit{et al.} \cite{shaik2025uncovering} & 5G SA Core (N2/N3/N4) & VPP/BESS Rules \& Depth Limit & \checkmark & $\times$ & $\times$ & \checkmark (6 Cores + SDR) \\
\midrule[1.5pt] 
Sattar \textit{et al.} \cite{sattar2025secure} & 5G Core Slicing & Proactive MILP Host Isolation & $\times$ & $\times$ & $\times$ & \checkmark (OpenVZ vEPC) \\
\rowcolor{blue!7}
Mahyoub \textit{et al.} \cite{mahyoub2025security} & 5G CP Procedures & Proactive MINLP VNF Sharing & $\times$ & $\times$ & $\times$ & $\times$ (Simulation) \\
Bolourian \textit{et al.} (\textit{5Guard}) \cite{bolourian20255guard} & E2E 5G Slicing & Online MIP Ensemble (L0--L2) & $\times$ & $\times$ & $\times$ & $\times$ (Simulation) \\
\rowcolor{blue!7}
AbdulGhaffar \& Matrawy \cite{aziz2026slice} & Multi-UPF Slicing & Proactive MINLP Phy/VNF Tiers & $\times$ (Topological) & $\times$ & $\times$ & \checkmark (Free5GC + Clab) \\
Khan \textit{et al.} (\textit{SliceSecure}) \cite{Khan10056693} & 5G Slice Data Plane & Reactive DL Anomaly Detection & $\times$ & $\times$ & $\times$ & \checkmark (Free5GC) \\
\midrule[1.5pt] 
\rowcolor{blue!7}
Parola \textit{et al.} \cite{parola2020proof} & 5G Mobile Gateway & eBPF/XDP Forwarding \& Policing & $\times$ & $\times$ (Policing Only) & $\times$ & \checkmark (Polycube PoC) \\
Messaoudi \textit{et al.} \cite{messaoudi2026brewing} & 5G UPF (N3/N6) & eBPF/XDP + \texttt{tc} HTB QoS & $\times$ & $\times$ (QoS Enforcement) & $\times$ & \checkmark (100G OAI-UPF) \\
\rowcolor{blue!7}
Akem \& Fiore \cite{akem2024towards} & 5G UPF Data Plane & In-Switch P4 Decision Tree ML & $\times$ & \checkmark (P4) & $\times$ & \checkmark (Tofino Switch) \\
Repetto \textit{et al.} (\textit{ASTRID}) \cite{repetto2022automating} & NFV / 5G UPF Edge & eBPF Sidecar + ARIMA Controller & $\times$ & $\sim$ ($\sim 2.4\text{ s}$ Load)  & $\times$ & \checkmark (Open5GS + K8s) \\
\rowcolor{blue!7}
Loureiro \textit{et al.} \cite{loureiro2024enhancing} & 5G Control Plane (NRF) & XDP Fixed-Window SYN Filter & $\times$ & \checkmark (TCP SYN Only) & $\times$ (Temp. Block) & \checkmark (Open5GS) \\
\midrule[1.5pt] 
\textbf{Proposed Defense} & \textbf{5G/6G Sliced UPF} & \textbf{Closed-Loop eBPF/XDP + UP-TA + CP-MO} & \textbf{\checkmark} & \textbf{\checkmark ($<0.4\ \mu\text{s}$)} & \textbf{\checkmark (SMF/AMF/UDR)} & \textbf{\checkmark (Free5GC + Clab)} \\
\bottomrule[1.5pt]
\end{tabular}
}
\end{table*}

\subsection{GTP-U Protocol Vulnerabilities and Encapsulation Attacks}
Designed over two decades ago for trusted telecom backhauls, the \gls{gprs} Tunneling Protocol User Plane (GTP-U) prioritizes packet forwarding based solely on \gls{ip} addresses and \gls{teid}, lacking native source authentication, integrity verification, or recursive payload inspection~\cite{shaik2025titanic}. Recent offensive security research has demonstrated that \gls{5g} \glspl{upf} inherit these fundamental protocol flaws~\cite{shaik2025uncovering}.

Industry threat intelligence from Trend Micro Research~\cite{salim2023attacks} first warned that compromised user devices in private \gls{5g} networks can inject malformed or nested \gls{gtp}-U payloads to trigger packet reflection and crash unpatched \glspl{upf}, recommending Cellular Technology (CT)-aware Deep Packet Inspection (DPI) firewalls for mitigation. At the global internet scale, Zhang~\textit{et al.}~\cite{zhang2025invade} scanned the publicly exposed \gls{gtp} nodes and identified around 749,000 exposed hosts across numerous providers and countries. The authors demonstrated that adversaries can exploit the vulnerabilities to perform various attacks. In contrast to our work, Zhang~\textit{et al.} focus on external Internet-facing \gls{gtp} exposure and suggest authentication-based protocol replacements rather than defending the internal N3 user plane against authenticated insider \glspl{ue} launching cross-slice nested \gls{gtp}-U attacks.

Focusing on the insider threats originating from authenticated \glspl{ue}, Kwon~\textit{et al.}~\cite{kwon2026mobile} demonstrated that encapsulating \gls{pfcp} (\gls{pfcp}-in-GTP) and ICMPv6-in-GTP payloads allows a malicious \gls{ue} to bypass \gls{dp} isolation and reach internal \gls{smf} and \gls{upf} nodes across operational \gls{5g} networks. However, while the authors discuss high-level routing and Access Control List (ACL) hardening recommendations, they do not design or empirically evaluate an implementable countermeasure to mitigate these attacks.

Most recently, Shaik \textit{et al.}~\cite{shaik2025uncovering, shaik2025titanic} evaluated Protocol Tunneling (PT) via nested \gls{gtp}-U-in-\gls{gtp}-U packets across four open-source and two commercial \gls{5g} cores. By exploiting predictable \gls{teid} allocations and permissive \gls{upf} routing, the authors executed multiple attacks, including direct \gls{ue}-to-\gls{ue} packet injection and reflective charging fraud. Although Shaik \textit{et al.} outline high-level hardening guidelines, they evaluate only unsliced cores and do not design or empirically evaluate an in-kernel defense or automated closed-loop session revocation framework.

\subsection{Proactive and Reactive 5G Network Slicing Defenses}
To prevent adversaries in one network slice from degrading co-hosted slices, existing slicing security literature largely follows two paradigms: proactive topological isolation via mathematical optimization and reactive anomaly detection via Machine Learning (ML).

In the proactive optimization domain, Sattar \textit{et al.}~\cite{sattar2025secure} formulated a \gls{milp} model that enforces hardware inter-slice isolation (ISI), intra-slice \gls{vnf} distribution, and dedicated secure hypervisor zones to mitigate \gls{ddos} flooding and slice-initiated CPU exhaustion attacks. Building on mathematical \gls{vnf} sharing, Mahyoub \textit{et al.}~\cite{mahyoub2025security} proposed a \gls{cp} procedure-level \gls{minlp} model that restricts \gls{vnf} sharing between \gls{3gpp} procedures using maximum \gls{vnf} traffic and external exposure constraints.

Bolourian \textit{et al.} \cite{bolourian20255guard} introduced \textit{5Guard}, an end-to-end (E2E) isolation-aware network slicing framework aimed at maximizing infrastructure provider profit. However, \textit{5Guard} evaluates resource overheads purely in numerical simulation without subjecting slices to live cyberattacks. A study by AbdulGhaffar and Matrawy~\cite{aziz2026slice} bridged mathematical optimization and empirical validation by proposing the \textit{Slice Isolation-as-a-Service} \gls{minlp} framework across multi-UPF data paths and deploying the generated topologies on a 10-\gls{vm} Free5GC and Nokia Containerlab testbed. Empirical evaluations under volumetric floods and cross-slice \gls{gtp}-in-\gls{gtp} reflection attacks proved that logical \gls{vnf}-only isolation on a shared physical host suffers significant performance degradation without physical hardware isolation. Unlike our runtime data-plane defense, these studies rely on an offline mathematical model to proactively isolate slices and reduce the attack surface.

On the other hand, reactive slicing defenses rely on ML and Deep Learning (DL) anomaly detectors, such as \textit{SliceSecure} by Khan \textit{et al.}~\cite{Khan10056693},  reinforcement learning-based \gls{ddos} slice isolation by Javadpour \textit{et al.}~\cite{Javadpour10064361}, and \gls{ran} performance isolation by Yarkina~\textit{et al.}~\cite{yarkina2022performance}. These reactive ML approaches evaluate flow-level features over conventional transport-layer flooding datasets that do not contain \gls{gtp}-in-\gls{gtp} protocol exploits, and their feature aggregation and inference delays are too slow to prevent immediate Linux kernel queue saturation during high-rate volumetric floods.

\subsection{Kernel-Level eBPF/XDP and Programmable UPF Defenses}
Recent works have also leveraged programmable \glspl{dp} (P4 and \gls{ebpf}/\gls{xdp}) within the \gls{5g} \gls{upf}. Parola~\textit{et al.}~\cite{parola2020proof} demonstrated an early proof-of-concept \gls{5g} mobile gateway in \gls{ebpf}/\gls{xdp} using the Polycube framework for traffic classification and \gls{qos} management. Extending this paradigm to carrier-grade \gls{3gpp} compliance, Messaoudi \textit{et al.}~\cite{messaoudi2026brewing} designed a high-performance OAI-UPF that combines \gls{xdp}, Linux Traffic Control (\texttt{tc-bpf}) and Hierarchical Token Bucket (HTB) queuing for \gls{qos} enforcement and traffic classification and forwarding. However, both Parola~\textit{et al.}~\cite{parola2020proof} and Messaoudi \textit{et al.}~\cite{messaoudi2026brewing} focus strictly on benign \gls{gtp} packets; their uplink \gls{xdp} parsers only strip a single outer \gls{gtp}-U header and neither inspect inner payloads for nested \gls{gtp}-in-\gls{gtp} exploits nor perform security mitigation.

Several studies have applied programmable data planes to \gls{5g} security and \gls{ddos} mitigation. Akem and Fiore~\cite{akem2024towards} embedded a trained tree-based ML classifier in a P4-programmable UPF to classify malicious \gls{5g} traffic at sub-microsecond line rates. Unlike our kernel-native \gls{ebpf}/\gls{xdp} solution that runs on commodity Linux \glspl{upf}, Akem and Fiore's solution requires specialized, expensive P4 switches and operates as an open-loop classifier without automated \gls{3gpp} control-plane session revocation. Repetto \textit{et al.}~\cite{repetto2022automating} developed an \gls{ebpf} framework to detect and mitigate \gls{dns} amplification attacks.

Loureiro~\textit{et al.}~\cite{loureiro2024enhancing} implemented an \gls{ebpf}/\gls{xdp} program to protect \gls{nrf} from \gls{tcp} SYN flood attacks in an Open5GS testbed. While effective at protecting \gls{nrf}, their solution only inspects unencapsulated transport-layer TCP SYN flags and temporarily throttles connections, rather than parsing N3 \gls{gtp}-U user-plane tunnels or permanently terminating compromised \gls{ue} sessions.

\textbf{Summary and Research Gap:} As pointed out in Table~\ref{tab:related_work}, a clear research gap exists in securing \gls{5g} and future \gls{6g} user planes against cross-slice encapsulation exploits. Offensive security studies~\cite{shaik2025uncovering,zhang2025invade,kwon2026mobile,salim2023attacks} expose the severity of nested \gls{gtp}-in-\gls{gtp} and protocol tunneling attacks. Proactive mathematical slice isolation models \cite{sattar2025secure, mahyoub2025security, bolourian20255guard, aziz2026slice} can prevent cross-slice bleeding only by provisioning dedicated physical servers. Meanwhile, existing \gls{ebpf}/\gls{xdp} and programmable \gls{upf} works either focus exclusively on benign forwarding ~\cite{messaoudi2026brewing, parola2020proof} or do not support the inspection of nested \gls{gtp}-U traffic~\cite{parola2020proof,messaoudi2026brewing, akem2024towards, repetto2022automating, loureiro2024enhancing}.

To the best of our knowledge, this work is the first to design, implement, and empirically validate a closed-loop \gls{ebpf}/\gls{xdp} security defense that unifies sub-microsecond ($<0.4\ \mu\text{s}$) nested \gls{gtp}-U packet dropping, kernel-level debounced telemetry, and asynchronous \gls{3gpp} control-plane session revocation and blacklisting of malicious \gls{ue} to preserve cross-slice \gls{qos} on shared \gls{upf} infrastructure.

\begin{figure*}[t]
    \centering
    \includegraphics[width=\linewidth]{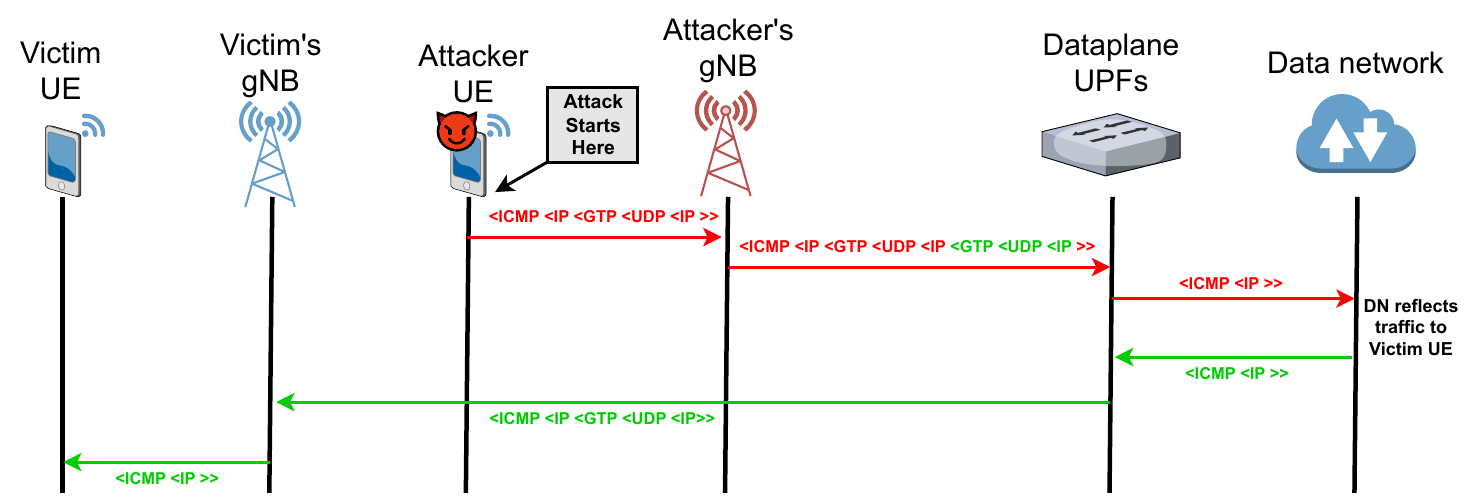}
    \caption{Reflection Attack Sequence Diagrams. \textcolor{red}{Red arrows for attack traffic.} \textcolor{Green}{Green is legitimate traffic, but becomes part of the attack}. \textbf{Note:} The attacker's gNB is not under the control of the attacker, but instead it is a legitimate gNB serving the attacker (malicious user)~\cite{shaik2025uncovering}. (Packet headers are shown in abbreviated inner-to-outer order to save space.)}
    \label{fig:reflection_call_flow}
\end{figure*}

\section{System Assumptions and Threat Model} \label{sec:threat_model}
To evaluate the proposed closed-loop defense, we define the threat landscape by explaining the system assumptions, adversary capabilities, and nested \gls{gtp} attack.

\subsection{System Assumptions}
\begin{itemize}
    \item \textbf{Infrastructure Security:} The \gls{5g} core network utilizes a virtualized architecture to enable multiple network slices deployment through shared or isolated \glspl{vnf}. The underlying bare-metal hardware and hypervisor configurations are assumed to be physically secure and tamper-proof~\cite{abdulghaffar2024impact, aziz2026slice}.
    \item \textbf{Control Plane Isolation:} The attack surface is strictly confined  to the \gls{dp}. \gls{cp} \glspl{nf}, such as \gls{smf} and \gls{amf}, are isolated and protected from attacks~\cite{aziz2026slice}.
    \item \textbf{Adversary Access:} The attacker is considered a legitimate \gls{ue} of the network; however, acting maliciously. The \gls{ue} successfully completes initial authentication and session establishment procedures to acquire a legitimate IP address and attach to a network slice~\cite{aziz2026slice,shaik2025uncovering}.
    \item \textbf{Trusted RAN and Source-IP Binding:} The serving \gls{gnb} is assumed to be a legitimate, uncompromised network node that encapsulates uplink \gls{ue} traffic with valid outer \gls{gtp}-U headers and \glspl{teid}. Furthermore, for the control-plane revocation proposed in this work, we assume that the compromised \gls{ue} transmits uplink packets using its legitimately assigned \gls{pdu} session \gls{ip} address in the non-nested inner \gls{ip} header, while spoofing victim identifiers inside the nested \gls{gtp}-U payload. The implications of relaxing this inner source-IP assumption and extending revocation to outer \gls{teid} correlation are discussed in Section~\ref{sec:limitation}.
\end{itemize}

\begin{figure*}[!t]
    \centering
    \includegraphics[width=\linewidth]{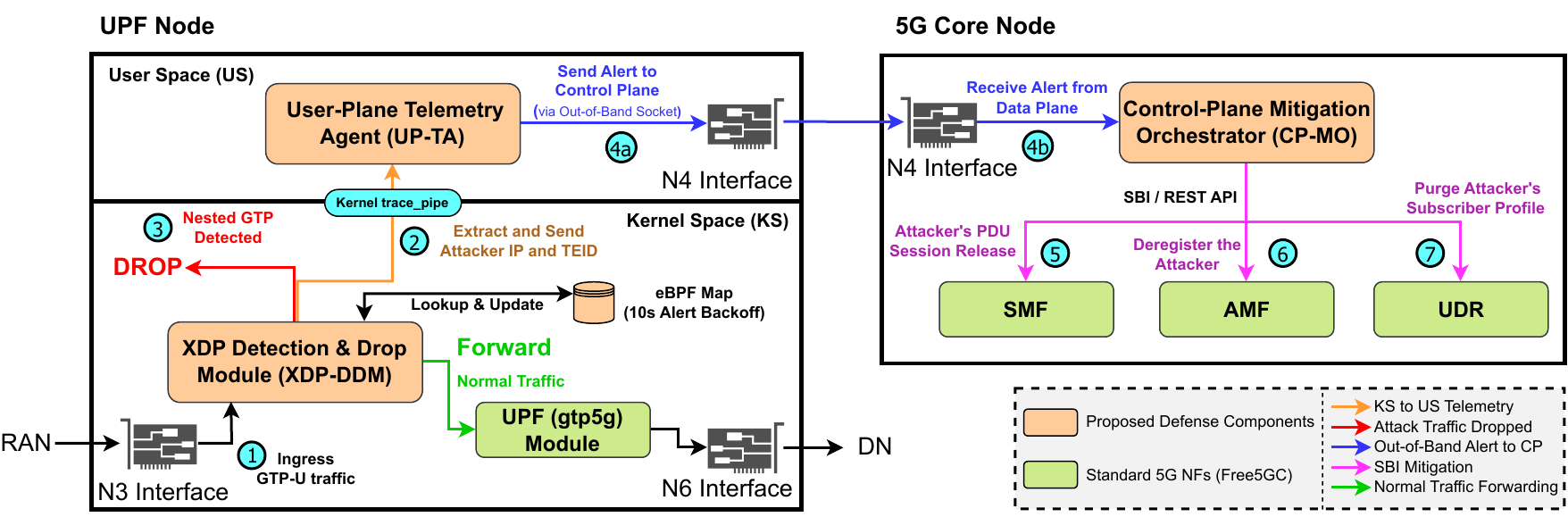}
    \caption{Proposed closed-loop eBPF/XDP detection and automated 5G/6G control-plane mitigation defense: (1) ingress N3 GTP-U inspection, (2) asynchronous telemetry emission via \texttt{trace\_pipe} governed by a 10-second eBPF backoff map, (3) inline line-rate packet dropping (\texttt{XDP\_DROP}), (4a--4b) out-of-band alert forwarding to the 5G Core, and (5--7) sequential SBI/REST session teardown, UE deregistration, and subscriber profile revocation across the SMF, AMF, and UDR, respectively. The green blocks represent Free5GC Network Functions (NFs)/components, whereas the orange blocks show our proposed closed-loop eBPF/XDP detection and automated mitigation components.}
    \label{fig:proposed_sol}
\end{figure*}

\subsection{Adversary Capabilities}
\begin{itemize}
    \item \textbf{Protocol Manipulation:} Possessing a comprehensive understanding of \gls{3gpp} specifications, the adversary can programmatically craft protocol-compliant \gls{ip} packets~\cite{aziz2026slice,shaik2025uncovering}.
    \item \textbf{Tunnel Injection:} The attacker injects customized payloads directly into the uplink data stream. The serving \gls{gnb} blindly encapsulates the malicious payload within legitimate \gls{gtp}-U headers and forwards it to the \gls{upf}~\cite{aziz2026slice,shaik2025uncovering}.
    \item \textbf{Reconnaissance and Spoofing:} Through probing, enumeration, or documentation, the attacker can identify the internal \gls{ip} addresses of critical core components. Leveraging this reconnaissance, the adversary spoofs internal \gls{ip} addresses and \glspl{teid} within the nested payloads to exploit \gls{upf} decapsulation logic~\cite{aziz2026slice,shaik2025uncovering}.
    \item \textbf{Volumetric Resource Exhaustion:} The attacker can generate flooding traffic from a single compromised \gls{ue}. By combining these volumetric floods with \gls{gtp}-U protocol tunneling, the attacker aims to disrupt the service of the victim \gls{ue} and to cause billing fraud~\cite{aziz2026slice}.
\end{itemize}

\subsection{Nested GTP Attack}
This attack was recently demonstrated in \gls{5g} networks by Shaik \textit{et al.} at ACM \gls{ccs} 2025~\cite{shaik2025uncovering}. Due to the lack of nested \gls{gtp} header verification and routing logic enforcement at the \gls{upf}, an attacker can exploit this vulnerability by injecting traffic into a victim's uplink \gls{gtp} tunnel~\cite{3gpp.29.281, mahyoub2024security}. The malicious actor creates an illegitimate nested \gls{gtp}-in-\gls{gtp} packet targeting the victim by setting the victim \gls{ue}'s \gls{ip} address and \gls{teid} in the spoofed inner \gls{ip} and \gls{gtp} headers. Furthermore, the attacker sets the inner destination \gls{ip} to the victim's serving \gls{upf}, and directs the inner payload (a \gls{dns} or \gls{icmp} request, etc.) to a public server.

The attacker then sends this crafted packet to their respective \gls{gnb}, which, as part of normal operation, adds the outer \gls{ip} header (with the serving \gls{gnb} and \gls{upf} N3 \gls{ip} addresses) and the outer \gls{gtp}-U header containing the attacker's uplink \gls{teid}, and forwards it to the \gls{upf}. Due to the lack of nested \gls{gtp} header verification at the \gls{upf}, it decapsulates the outer \gls{gtp} headers and routes the inner payload based on the inner malformed \gls{gtp} header. As the attacker crafted the inner headers with the information of the victim, the \gls{upf} routes the packet through the victim's uplink tunnel, appearing to be originated from the victim \gls{ue}. This, in turn, results in routing the public server's response towards the victim \gls{ue} through its downlink tunnel. As a consequence of this attack, the victim's uplink and downlink tunnels are filled with attack traffic, causing quota depletion, billing fraud, and reflection attacks. Figure~\ref{fig:reflection_call_flow} shows the sequence of the nested \gls{gtp} attacks.

\section{Proposed Closed-Loop eBPF Defense} \label{sec:proposed}
\subsection{Overview and Design Principles} \label{sec:design}

To neutralize the threat of nested \gls{gtp}-U attacks without compromising the performance of the legitimate user traffic, we propose a closed-loop detection and mitigation defense. This solution operates across the \gls{5g} \gls{dp} and \gls{cp} boundaries, leveraging \gls{ebpf}.

\gls{ebpf} was selected as the foundational technology for this defense due to its unique capabilities in modern cloud-native environments. \gls{ebpf} operates as an abstract virtual machine that executes user-defined programs inside a secure sandbox within the Linux kernel, allowing developers to extend kernel behavior without modifying the kernel source code or risking system crashes~\cite{he2023cross}. To achieve the stringent low-latency requirements of \gls{5g}/\gls{6g} data planes, this defense specifically relies on \gls{xdp}, which is a high-performance, programmable network data path integrated directly into the Linux kernel~\cite{hoiland2018express}. It provides a safe execution environment for custom packet processing applications within the network device driver context~\cite{hoiland2018express}. By shifting data processing to this earliest possible point, before the kernel allocates memory or engages the standard networking stack, \gls{xdp} enables wire-speed packet manipulation and dropping~\cite{hoiland2018express, soldani2023ebpf}.

Leveraging these capabilities, the proposed defense is built upon three core design principles:
\begin{enumerate}
    \item \textbf{Lightweight Processing:} Utilizing a minimal footprint to guarantee low-latency packet processing at wire speed.
    \item \textbf{High Reliability:} Ensuring rapid and accurate detection of malicious nested \gls{gtp}-U traffic.
    \item \textbf{Plug-and-Play Integration:} Designed to be fully compatible with any Linux-based \gls{upf} deployment in current \gls{5g} or future \gls{6g} networks.
\end{enumerate}

\begin{figure*}
    \centering
    \includegraphics[width=\linewidth]{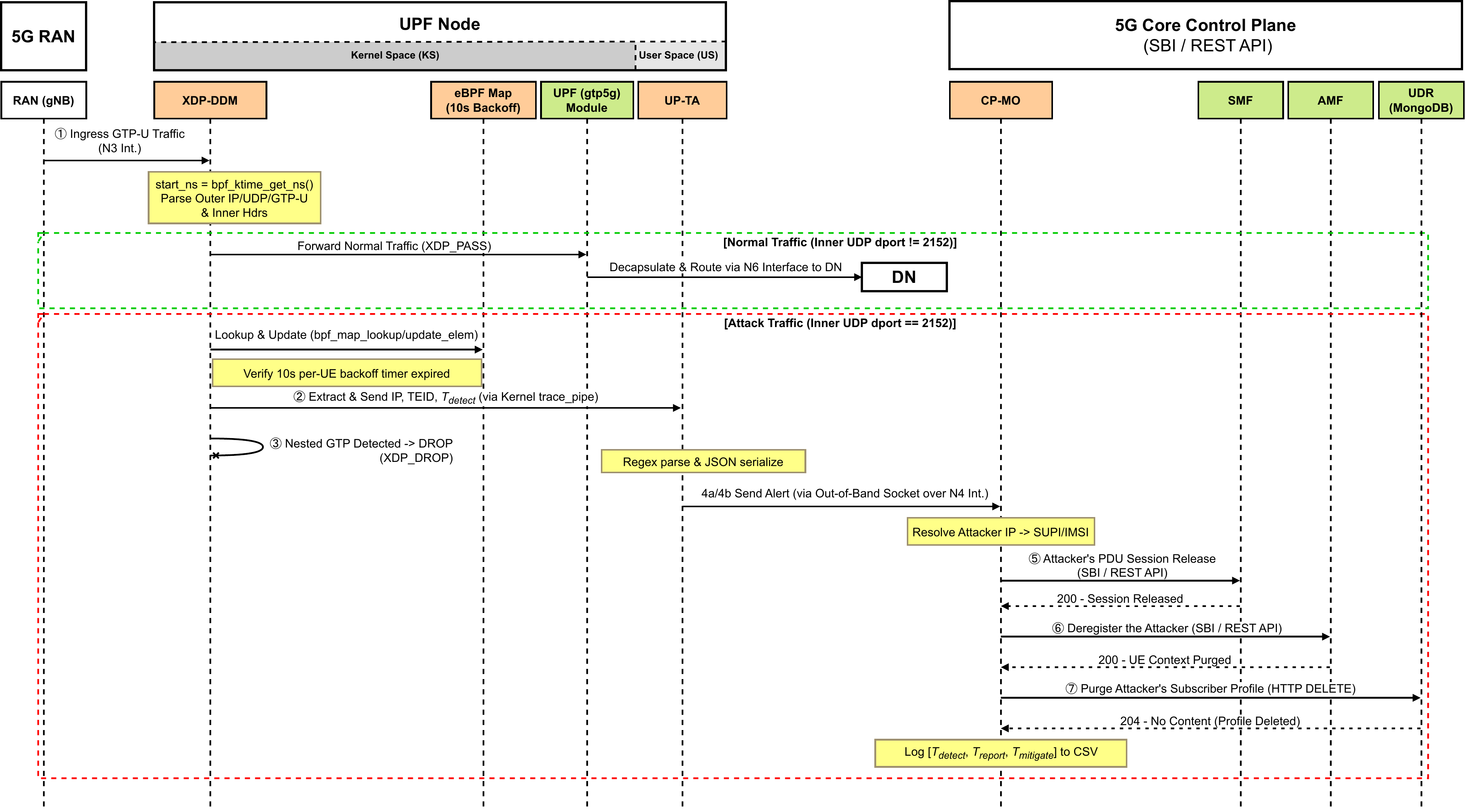}
    \caption{Call Flow of the Proposed Closed-Loop eBPF Defense. The Green dotted block shows the flow for legitimate traffic, while the Red dotted block shows the procedure for nested \gls{gtp} attack traffic.}
    \label{fig:call_flow}
\end{figure*}

As illustrated in Figure~\ref{fig:proposed_sol} and the call flow in Figure~\ref{fig:call_flow}, the proposed defense consists of three specialized components: \acrfull{xdp-ddm}, \acrfull{up-ta}, and the \acrfull{cp-mo}.

\subsection{XDP Detection \& Drop Module (XDP-DDM)} \label{sec:xdp_ddm}
The \gls{xdp-ddm} is a lightweight, kernel-space enforcement module attached to the \gls{upf}'s N3 ingress interface (Algorithm \ref{alg:xdp_ddm}). Operating at the \gls{xdp} hook, it intercepts traffic at the earliest possible stage in the networking stack.

To maintain the highest processing speeds, the \gls{xdp-ddm} restricts its operations to purely line-rate header parsing and packet dropping. \textcircled{1} \gls{xdp-ddm} intercepts the arriving frames from the \gls{ran} and inspects the outer \gls{gtp}-U headers and parses the encapsulated payload. Legitimate user-plane packets immediately return \texttt{XDP\_PASS} and are forwarded to the \texttt{\gls{upf} (gtp5g) Module}~\cite{go_upf,gtp5g} for standard decapsulation toward the \gls{dn} via the N6 Interface.  \textcircled{2} If a nested \gls{udp} packet destined for port 2152 is detected, the \gls{xdp-ddm} queries the \texttt{\gls{ebpf} Map} indexed by the attacker \gls{ip} to enforce a $10$-second alert backoff timer. This prevents exhausting CPU resources during a volumetric flooding attack. Rather than executing computationally expensive JSON serialization in the kernel, the module asynchronously sends raw telemetry (Attacker \gls{ip}, \gls{teid}, monotonic timestamp, and \gls{xdp} detection delay ($T_{detect}$)) to user-space via the kernel \texttt{trace\_pipe}. \textcircled{3} Immediately following the transmission of telemetry (or detecting the attack traffic), the \gls{xdp-ddm} returns \texttt{XDP\_DROP}, discarding the malicious frame at line rate.

\begin{algorithm}[t]
\caption{XDP-DDM: XDP Detection \& Drop Module (UPF Kernel-Space)}
\label{alg:xdp_ddm}
\SetAlgoLined
\KwIn{Ingress packet $p$}
\KwOut{Action (\texttt{XDP\_PASS} or \texttt{XDP\_DROP})}
\SetKwInput{KwState}{Shared State}
\KwState{$\mathcal{M}_{state} \leftarrow$ BPF LRU Hash Map (Debounce tracking)}
\KwState{$\tau_{db} \leftarrow 10$ seconds (Debounce timeout)}
\BlankLine

$t_{start} \leftarrow$ current nanosecond\;
\If{$p$ lacks valid outer IPv4/UDP/GTP headers}{
    \Return \texttt{XDP\_PASS} \tcp*{Step 1: Parse Outer Headers}
}

$teid \leftarrow p.gtp.teid$ \tcp*{Extract Outer TEID}
Extract encapsulated payload as $p_{inner}$\;

\tcp{Log Normal Traffic Execution Time (Evaluation Baseline)}
\If{$p_{inner}.protocol == \text{ICMP}$}{
    $T_{detect} \leftarrow \text{current nanosecond} - t_{start}$\;
    Log $T_{detect}$ to trace\_pipe\;
    \Return \texttt{XDP\_PASS}\;
}

\If{$p_{inner}.protocol \neq \text{UDP}$}{
    \Return \texttt{XDP\_PASS} \tcp*{Pass non-UDP payloads}
}

\tcp{Step 2: Detect Nested GTP-U Signature}
\If{$p_{inner}.dest\_port == 2152$}{
    $t_{end} \leftarrow$ current nanosecond\;
    $T_{detect} \leftarrow t_{end} - t_{start}$\;
    $ip_{src} \leftarrow p_{inner}.saddr$\;
    
    $t_{last} \leftarrow \mathcal{M}_{state}[ip_{src}]$\;
    \If{$t_{last}$ is NULL \textbf{or} $(t_{end} - t_{last}) > \tau_{db}$}{
        $\mathcal{M}_{state}[ip_{src}] \leftarrow t_{end}$\;
        Export $\langle ip_{src}, teid, t_{end}, T_{detect} \rangle$ to User-Space (UP-TA)\;
    }
    \Return \texttt{XDP\_DROP} \tcp*{Step 3: Drop \& Export Telemetry}
}
\BlankLine
\Return \texttt{XDP\_PASS} \tcp*{Pass legitimate inner UDP traffic}
\end{algorithm}

\subsection{User-Plane Telemetry Agent (UP-TA)} \label{sec:up_ta}
The \gls{up-ta} acts as the intermediary bridging the \gls{dp} line-rate and the core \gls{cp} (Algorithm~\ref{alg:up_ta}). Deployed as a user-space daemon on the \gls{upf} node, it continuously polls the \gls{ebpf} \texttt{trace\_pipe} for incoming threat telemetry.

Upon receiving an event, the \gls{up-ta} calculates the kernel-to-user reporting delay and formats the raw parameters into a structured JSON alert. This decoupling ensures that the \gls{dp}'s dropping efficiency remains unaffected by telemetry serialization. \textcircled{4} Finally, the \gls{up-ta} dispatches the JSON alert to the \gls{cp-mo} via an out-of-band \gls{udp} socket established over the N4 interface.

\begin{algorithm}[t]
\caption{UP-TA: User-Plane Telemetry Agent (UPF User-Space)}
\label{alg:up_ta}
\SetAlgoLined
\KwIn{Event stream from BPF TracePipe}
\BlankLine
\While{listening to BPF TracePipe}{
    On receiving event $e = \langle ip_{src}, teid, t_{end}, T_{detect} \rangle$\;
    $t_{arr} \leftarrow$ current nanosecond\;
    $\Delta t_{hop1} \leftarrow t_{arr} - t_{end}$ \tcp*{Hop 1: Kernel-to-User delay}
    
    $\mathcal{A} \leftarrow$ construct JSON Alert $\langle ip_{src}, teid, \Delta t_{hop1}, T_{detect} \rangle$\;
    Forward Alert $\mathcal{A}$ to CP-MO via N4 \tcp*{Step 4: Dispatch Telemetry}
}
\end{algorithm}

\subsection{Control-Plane Mitigation Orchestrator (CP-MO)} \label{sec:cp_mo}
The \gls{cp-mo} serves as the final component of the closed-loop defense, responsible for permanently revoking the attacker's network access (Algorithm~\ref{alg:cp_mo}). Residing within the \gls{5g} \gls{cp}, it continuously listens for incoming \gls{udp} alerts from the \gls{up-ta}.

To prevent socket blocking during coordinated attacks, the \gls{cp-mo} utilizes an asynchronous, multi-threaded execution mode. Upon resolving the attacker's \gls{ip} to a specific \gls{supi} and SM (Session Management) Context, it triggers a sequential, three-stage mitigation via \gls{sba} REST \glspl{api}. \textcircled{5} First, it issues a release command to the \gls{smf} to release the attacker's \gls{pdu} session and delete the \gls{upf} tunnels. \textcircled{6} Secondly, it commands the \gls{amf} to purge the signaling context and deregister the attacker \gls{ue}. \textcircled{7} Finally, it invokes an HTTP DELETE request to the \gls{udr}, permanently blacklisting the attacker's subscriber profile based on their permanent identifier (\gls{supi}). At the end of the closed-loop mitigation, \gls{cp-mo} logs the \gls{xdp-ddm} detection delay ($T_{detect}$), reporting delay ($T_{report}$), and \gls{cp-mo} mitigation delay ($T_{mitigate}$).

\begin{algorithm}[t]
\caption{CP-MO: Control-Plane Mitigation Orchestrator (Control Plane)}
\label{alg:cp_mo}
\SetAlgoLined
\KwIn{Alert $\mathcal{A} = \langle ip_{src}, teid, \Delta t_{hop1}, T_{detect} \rangle$ from UP-TA}
\SetKwProg{Proc}{Procedure}{}{}
\BlankLine

\While{listening on UDP port}{
    Receive Alert $\mathcal{A}$\;
    $t_{n4\_arr} \leftarrow$ current nanosecond\;
    Parse JSON payload $\mathcal{A}$\;
    
    \tcp{Calculate Total N3-to-N4 Reporting Time}
    $T_{report} \leftarrow \mathcal{A}.\Delta t_{hop1} + (\text{current nanosecond} - t_{n4\_arr})$\;
    
    \tcp{Asynchronous execution to maintain non-blocking socket}
    \textbf{Spawn Background Thread} \texttt{Mitigate}($\mathcal{A}.ip_{src}, \mathcal{A}.teid, \mathcal{A}.T_{detect}, T_{report}$)\;
}
\BlankLine

\Proc{\texttt{Mitigate}($ip_{src}, teid, T_{detect}, T_{report}$)}{
    $t_{mit\_start} \leftarrow$ current nanosecond\;
    $\mathcal{L}_{ue} \leftarrow$ HTTP GET AMF Registered UEs\;
    $supi, sm\_ctx, plmn \leftarrow$ NULL, NULL, NULL\;
    
    \ForEach{$ue \in \mathcal{L}_{ue}$}{
        \If{$ue$ contains $ip_{src}$}{
            $supi \leftarrow ue.SUPI$\;
            $sm\_ctx \leftarrow ue.SmContextRef$\;
            $plmn \leftarrow ue.PLMN$\;
            \textbf{break} \tcp*{Resolve Attacker SUPI \& SM Context}
        }
    }
    
    \If{$supi \neq$ NULL \textbf{and} $sm\_ctx \neq$ NULL}{
        \tcp{Step 5: Trigger UPF Session Teardown}
        HTTP POST SMF($sm\_ctx$, release) 
        \tcp*{Step 6: Trigger Deregistration}
        HTTP POST AMF($supi$, release)\;
        \tcp{Step 7: Blacklist UE}
        HTTP DELETE UDR($supi,~plmn$)\;
    }
    
    $T_{mitigate} \leftarrow \text{current nanosecond} - t_{mit\_start}$\;
    Append $\langle ip_{src}, teid, T_{detect}, T_{report}, T_{mitigate} \rangle$ to Evaluation Logs\;
}
\end{algorithm}

\subsection{Implementation Details and State Machine Synchronization} 
Translating this defense into a functioning \gls{5g} prototype required overcoming significant \gls{sba} state-synchronization challenges. By default, when the Free5GC~\cite{free5gc} \gls{amf} triggers a Network-Initiated Deregistration, it only deletes its local context but ignores the active \gls{dp} \gls{pdu} tunnels allocated to the \gls{ue}. To prevent ghost tunnels and ensure the \gls{amf} and \gls{smf} state machines remain perfectly synchronized per \gls{3gpp} TS 23.502~\cite{3gpp.23.502}, our \gls{cp-mo} orchestrates a strict teardown sequence. The system explicitly invokes the \texttt{Nsmf\_PDUSession\_ReleaseSMContext} service operation for every active \gls{pdu} session tied to the attacker, forcing the \gls{smf} to tear down the \gls{upf} tunnels before the \gls{ue} is fully deregistered and blocked. While session release and deregistration invoke \gls{3gpp} SBI operations, \gls{ue} \gls{ip}-to-\gls{supi} resolution and \gls{udr} profile removal are executed via Free5GC's WebConsole management REST \gls{api}.

While recent open-source community prototypes have demonstrated the basic feasibility of using \gls{ebpf}/\gls{xdp} for unencapsulated application-layer filtering (such as \gls{dns}) on \gls{upf} interfaces~\cite{free5gc2026dpi5g}, our proposed defense utilizes \gls{ebpf}/\gls{xdp} for a comprehensive security framework.  Specifically, the design and scope of our work fundamentally differ by parsing recursive \gls{gtp}-in-\gls{gtp} encapsulation, implementing kernel-space debouncing, and closing the loop with automated \gls{cp} revocation.

\section{Evaluation Methodology} \label{sec:methodology}
This section outlines the experimental methodology utilized to evaluate the efficacy, resilience, and performance impact of the proposed closed-loop \gls{ebpf} defense.

\subsection{Testbed Environment}
\begin{figure}[!t]
    \centering
    \includegraphics[width=\linewidth]{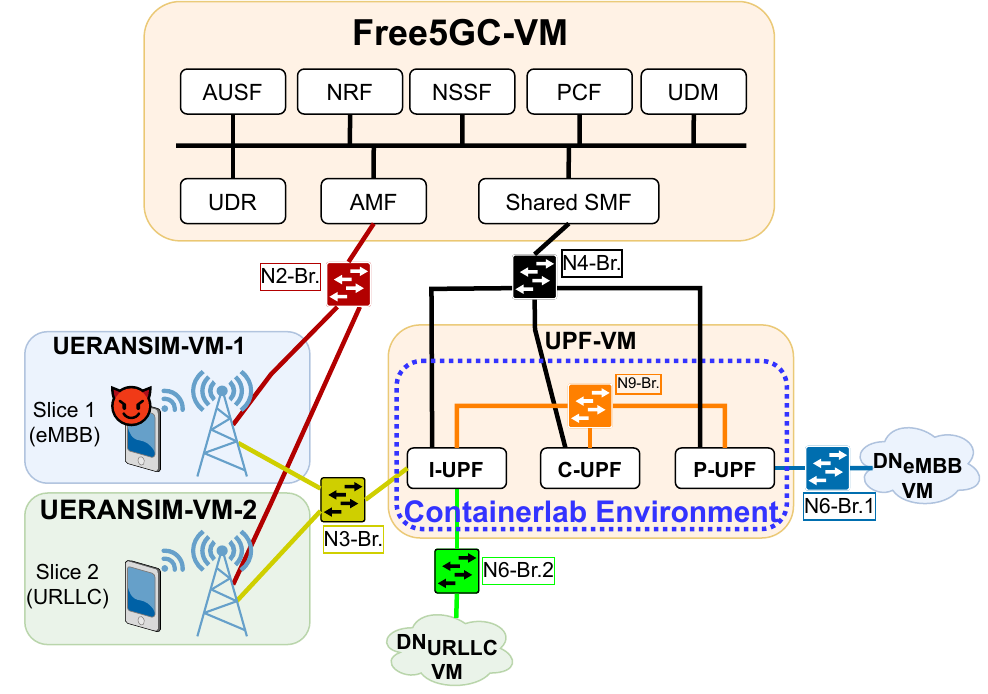}
    \caption{Experimental 5G network slicing testbed architecture. The control plane relies on Free5GC, while the RAN and UEs are emulated via UERANSIM across dedicated VMs for the eMBB (Slice 1) and URLLC (Slice 2) slices. The data plane is orchestrated using Containerlab, featuring a multi-tiered UPF topology interconnected via dedicated bridges (N2, N3, N4, N6, and N9) to isolate traffic and validate the eBPF defense under cross-slice attack scenarios.}
    \label{fig:testbed}
\end{figure}

To rigorously evaluate the proposed closed-loop \gls{ebpf} defense, we utilize the \gls{5g} network slicing testbed, as shown in Figure~\ref{fig:testbed}. The \gls{5g} Core is deployed using Free5GC~\cite{free5gc} in a dedicated \gls{vm}, providing the requisite \glspl{nf}, while the \gls{ran} and \gls{ue} are emulated using UERANSIM~\cite{ueransim}. Two distinct \glspl{vm} are used to host the emulated \gls{ue} and \gls{gnb} instances for Slice 1 (\gls{embb}) and Slice 2 (\gls{urllc}). Each slice is served by its corresponding \gls{dn}: $\gls{dn}_{eMBB}$ and $\gls{dn}_{URLLC}$, respectively. The \gls{5g} \gls{dp} \glspl{upf} are deployed and orchestrated using Containerlab~\cite{containerlab} inside a VirtualBox \gls{vm}~\cite{virtualbox} to allow flexible and custom data path creation. This containerized topology provides a highly reproducible environment for deploying our \gls{xdp} hooks and executing high-fidelity traffic generation.

To accurately model a realistic core network, we assume a multi-tiered \gls{upf} topology consisting of intermediate \glspl{upf} ($I-\gls{upf}$), Central \glspl{upf} ($C-\gls{upf}$), and \gls{pdu} Session Anchor \glspl{upf} ($P-\gls{upf}$). The $I-\gls{upf}$ acts as an edge branching node, routing latency-critical \gls{urllc} traffic directly to an edge \gls{dn} while forwarding \gls{embb} traffic deeper into the core. The $C-\gls{upf}$ acts purely as a transit node to mimic multi-hop core transit and does not connect to any \gls{dn}. Finally, the $P-\gls{upf}$ terminates the connection to the cloud \gls{dn}. Consequently, the \gls{pdu} data path for \gls{urllc} is modeled as ($\gls{gnb}  \rightarrow I-\gls{upf} \rightarrow \gls{dn}_{URLLC}$), whereas the path for \gls{embb} slices is modeled as ($\gls{gnb} \rightarrow I-\gls{upf} \rightarrow C-\gls{upf} \rightarrow P-\gls{upf} \rightarrow \gls{dn}_{eMBB}$). 

\subsection{Resilience Under Volumetric Attack} 
To evaluate the resilience of our defense under active attack, we orchestrate a cross-slice nested \gls{gtp} volumetric flood. The attacker \gls{ue} is provisioned in Slice 1, while the victim \gls{ue} operates in Slice 2. To circumvent the inherent limitation of the UERANSIM software crashing under high traffic generation scenarios~\cite{uransim_error}, we utilize the TRex traffic generator~\cite{trex}. A crafted nested \gls{gtp} payload is captured into a PCAP file and aggressively replayed by TRex at a sustained rate of 10,000 \gls{pps}.

Throughout the attack window, we continuously measure the victim \gls{ue}'s \gls{rtt} delay and \gls{tcp} throughput. Simultaneously, the penetrating attack traffic is captured via the victim \gls{ue}'s UERANSIM tunnel and logged to a PCAP file. This methodology enables a direct comparison of network resilience between the baseline environment (without the proposed \gls{ebpf}/\gls{xdp} defense) and the mitigation environment.

\subsection{Performance Impact on Legitimate Traffic}
Carrier-grade \gls{5g} deployments demand that security mechanisms do not degrade the \gls{qos} for benign users. To quantify the system overhead, we measure the impact of the proposed \gls{ebpf}/\gls{xdp} program on legitimate user traffic.

The evaluation focuses on two primary performance metrics: \textbf{End-to-End \gls{rtt} Delay} and \textbf{\gls{tcp} Throughput}. We conduct clean tests using standard \gls{icmp} and \gls{tcp} flows traversing \gls{urllc} and \gls{embb} slices in an attack-free environment.  By recording these metrics, we establish a baseline and compare it directly against the user's end-to-end delay and throughput when the \gls{ebpf}/\gls{xdp} program is enabled at the \gls{upf}.

\subsection{Delay of the Closed-Loop Defense} \label{sec:Mitigation_eff}
To precisely quantify the responsiveness of the closed-loop pipeline, we decompose the total system delay into three distinct metrics corresponding to the architectural components:
\begin{itemize}
    \item  \textbf{Detection Delay ($T_{detect}$):} The execution time required for the \gls{xdp-ddm} to identify an ingress packet as either normal or malicious traffic. It is measured from the moment the \gls{ebpf}/\gls{xdp} program receives the packet until the forwarding or dropping decision is finalized. For attack traffic, this timing mechanism is natively integrated into the signature detection logic. To establish a baseline for normal traffic, the \gls{ebpf} program was specifically instrumented to measure the execution time of benign \gls{icmp} packets during the evaluation.
    \item \textbf{Reporting Delay ($T_{report}$):} The time elapsed to dispatch a threat alert to the \gls{5g} core network. This is computed as the sum of the kernel-to-user transfer delay (from \gls{xdp-ddm} detection to \gls{up-ta} reception via \texttt{trace\_pipe}) and the JSON payload parsing time at the \gls{cp-mo}, as detailed in Algorithm~\ref{alg:up_ta} and Algorithm~\ref{alg:cp_mo}. As noted in the Limitations (Section~\ref{sec:limitation}), this measurement explicitly excludes the virtual switch delay introduced by inherent hypervisor clock skew.
    \item \textbf{Mitigation Delay ($T_{mitigate}$):} The duration required for the \gls{cp-mo} to successfully orchestrate the mitigation. This metric wraps the execution of the control-plane \texttt{Mitigate} procedure (\texttt{execute\_mitigation}). 
\end{itemize}

\begin{figure}[t]
    \centering
    \includegraphics[width=\linewidth]{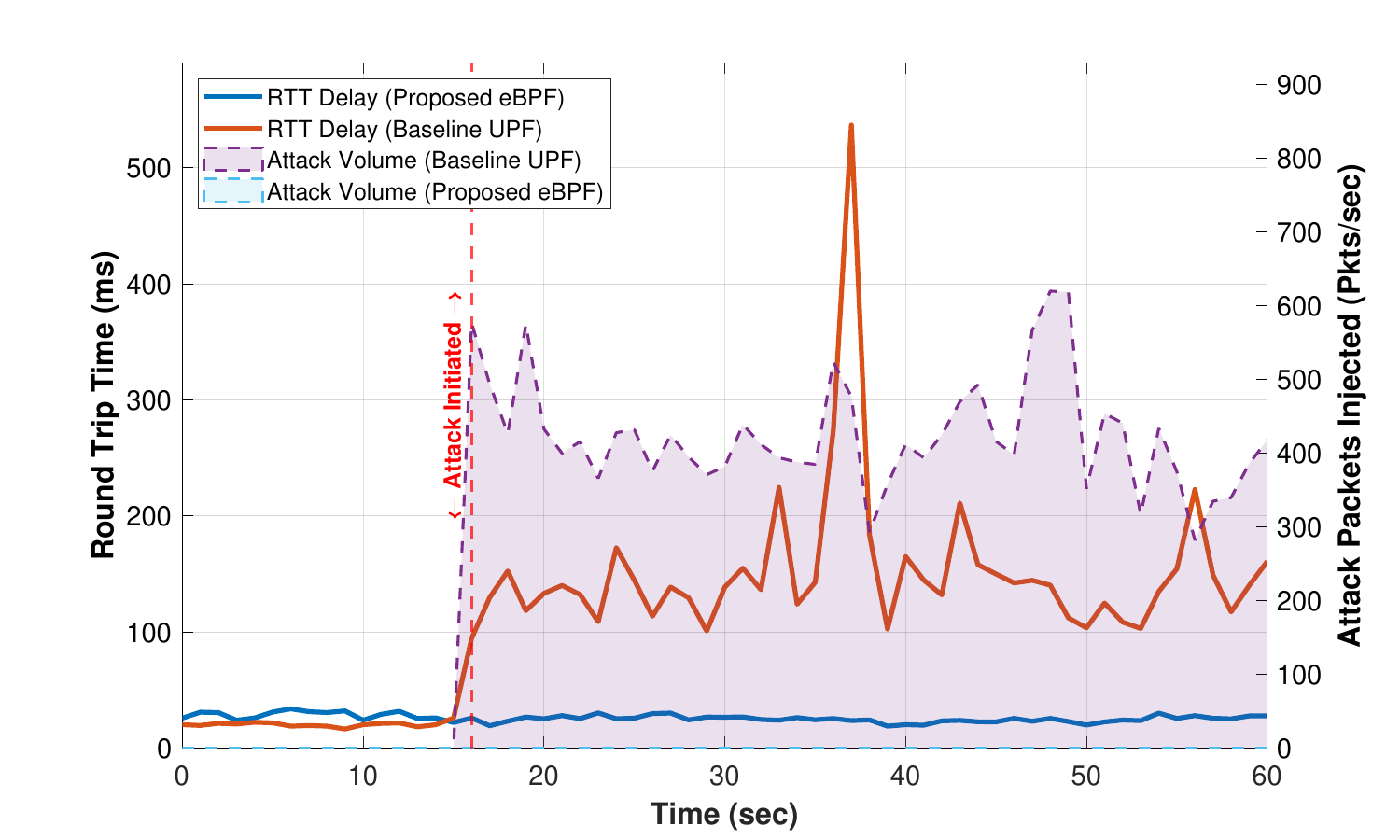}
    \caption{Dual-axis evaluation of Round-Trip Time (RTT) resilience for a victim UE in Slice 2. The baseline UPF suffers severe impact and latency spikes ($>500$ ms) during the nested GTP-U flood, while the proposed eBPF defense successfully shields the victim, maintaining a stable baseline delay.}
    \label{fig:rtt}
\end{figure}

\begin{figure}[t]
    \centering
    \includegraphics[width=\linewidth]{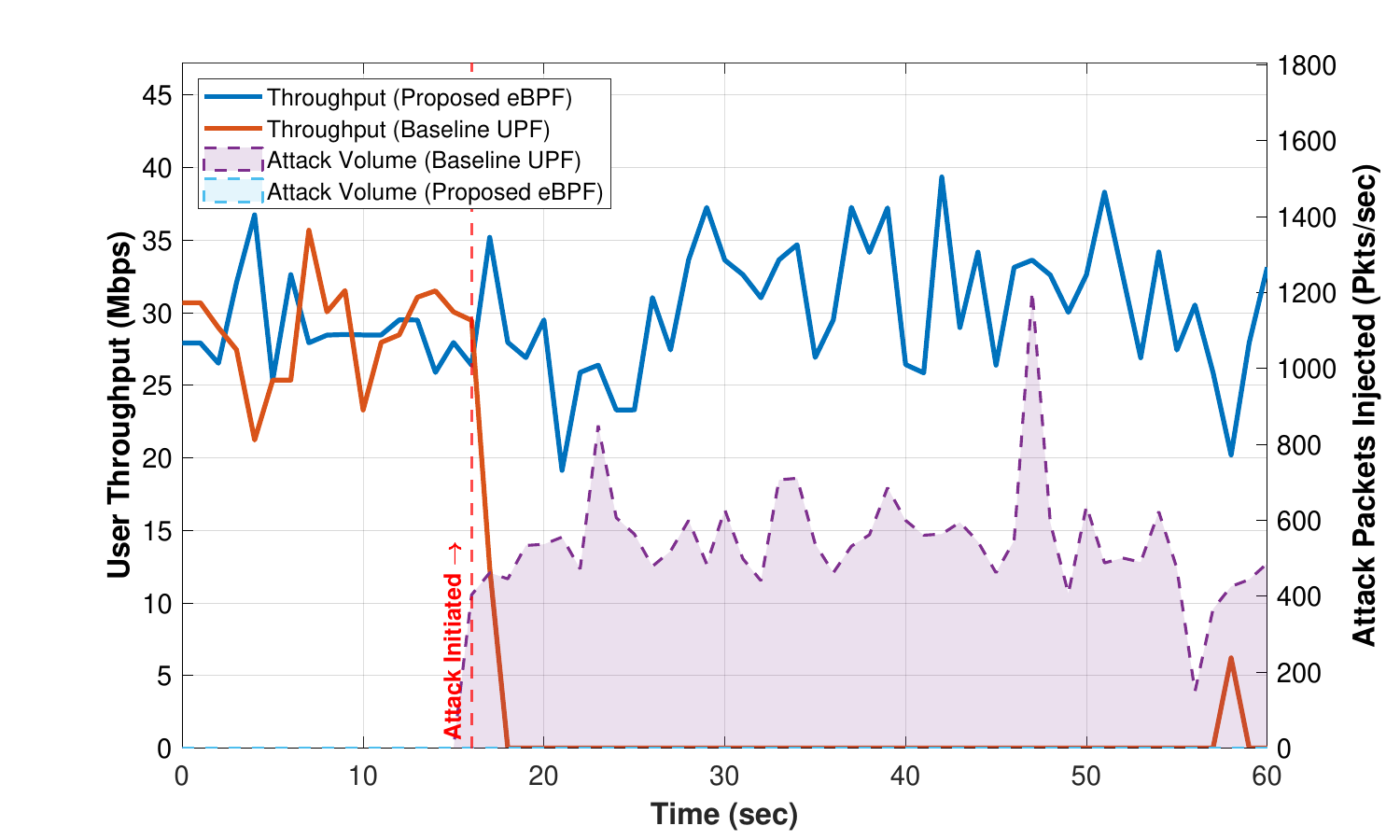}
    \caption{Dual-axis evaluation of TCP throughput resilience for a victim UE in Slice 2. The baseline UPF experiences complete resource exhaustion (0 Mbps) as attack traffic floods the queues. In contrast, the proposed eBPF defense perfectly isolates the slices, silently dropping the attack volume and preserving an optimal $25-40$ Mbps throughput for the victim.}
    \label{fig:throughput}
\end{figure}

\section{Experimental Results} \label{sec:results}

\subsection{Resilience Under Volumetric Attack}  \label{sec_result_resilience}
Figure~\ref{fig:rtt} and Figure~\ref{fig:throughput} illustrate the resilience of the victim \gls{ue} deployed in \gls{urllc} Slice 2 against a nested \gls{gtp}-U volumetric flood originating from an attacker in Slice 1. Both figures utilize a dual-axis representation, mapping the primary user performance metrics, including \gls{rtt} and Throughput, against the volume of injected attack packets penetrating the victim's tunnel.

\textbf{\textit{Pre-Attack Baseline Stability:}} During the initial pre-attack duration (0 to 15 seconds), the \gls{rtt} delay and \gls{tcp} throughput are highly comparable in both scenarios. The baseline \gls{upf} and the proposed \gls{ebpf} solution exhibit nearly identical performance, establishing that the \gls{ebpf} hook introduces minimal overhead to legitimate user traffic under normal operating conditions.

\textbf{\textit{Severe Impact on the Baseline UPF:}} At $t=16$~s, the nested \gls{gtp} volumetric attack is initiated (shown in the pink shaded area). In the baseline \gls{upf} scenario (without the proposed \gls{ebpf} solution), the injected nested \gls{gtp} attack traffic instantly overwhelms the core. The victim's \gls{rtt} increases significantly, remaining above $100$ ms throughout the attack duration and peaking at over $500$ ms in the worst-case scenario, as shown in Figure~\ref{fig:rtt}. On the other hand, the victim's throughput drops entirely to zero during the attack, as illustrated in Figure~\ref{fig:throughput}, causing a massive degradation in \gls{qos} and \gls{qoe}.

Although the attacker injects traffic at a rate of 10,000 \gls{pps} at the source, the attack volume traversing the victim's tunnel registers between 500 and 1,200 \gls{pps}. This discrepancy highlights the core mechanism of the \gls{dos} attack: the Linux kernel's standard \texttt{gtp5g} module becomes entirely saturated. As the queues reach maximum capacity, the operating system indiscriminately drops the majority of incoming packets, effectively starving the victim's legitimate traffic. The brief 6 Mbps throughput anomaly observed at $t=58$ seconds is a characteristic artifact of \gls{tcp} starvation, where network window probes temporarily succeed during micro-fluctuations in the congested queue.

\textbf{\textit{Resilience with Proposed Defense:}} In stark contrast, when the proposed \gls{ebpf} defense is enabled, the user's \gls{rtt} delay and \gls{tcp} throughput remain completely consistent with pre-attack levels, as demonstrated in Figure~\ref{fig:rtt} and Figure~\ref{fig:throughput}. The attack volume recorded in the victim's tunnel remains at absolute zero, indicating that the malicious nested \gls{gtp} traffic is intercepted and dropped immediately by the \gls{xdp-ddm} at the network driver level. While the line-rate \gls{xdp-ddm} absorbs the attack shock, the \gls{cp-mo} asynchronously revokes the attacker's connection and blacklists the attacker \gls{ue} to prevent future associations (as quantified in Section~\ref{sec_result_latency}).

\begin{tcolorbox}[
    colback=blue!7,   
    colframe=blue!50!black,  
    boxrule=0.5pt,           
    arc=3pt,                 
    left=2pt, right=2pt, top=2pt, bottom=2pt 
]
\textbf{\textit{Key Takeaway:}} The proposed \gls{ebpf} defense successfully enforces strict cross-slice isolation at the data plane. While standard \gls{upf} suffers severe connection deterioration due to queue saturation during nested \gls{gtp} volumetric floods, our line-rate \gls{xdp-ddm} guarantees minimal overhead in the victim's legitimate traffic, maintaining optimal \gls{qos}.
\end{tcolorbox}

\begin{figure}[t]
    \centering
    \includegraphics[width=\linewidth]{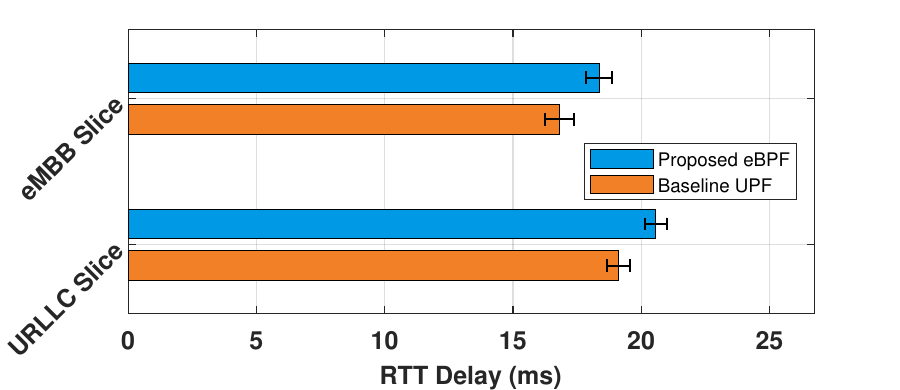}
    \caption{Impact of the proposed eBPF defense on legitimate end-to-end RTT delay across eMBB and URLLC slices over 100 samples (95\% CI), demonstrating minimal latency overhead ($<2\text{ ms}$) compared to the baseline UPF.}
    \label{fig:rtt_overhead}
\end{figure}

\subsection{Performance Impact on Legitimate Traffic}  \label{sec_result_impact}
Carrier-grade security solutions must operate transparently without penalizing benign users. To quantify the operational overhead of the proposed defense under normal, attack-free conditions, we evaluate its impact on legitimate \gls{ue} traffic across both the \gls{embb} and \gls{urllc} network slices.

\subsubsection{Impact on End-to-End Delay (RTT)}
Figure~\ref{fig:rtt_overhead} illustrates the end-to-end \gls{rtt} delay for legitimate \gls{icmp} traffic traversing the \gls{embb} and \gls{urllc} slices, comparing the baseline \gls{upf} against the \gls{upf} running the proposed \gls{ebpf} solution. To ensure statistical rigor, each bar represents the mean \gls{rtt} computed across 100 clean packet samples, accompanied by its $95\%$ confidence interval. 

In the baseline configuration, the \gls{embb} slice exhibits an average \gls{rtt} of  $16.8~\text{ms}$, while the \gls{urllc} slice averages $19.1$ ms. Although the \gls{embb} topology traverses three \gls{upf} hops compared to a single \gls{upf} hop in the \gls{urllc} slice, the \gls{urllc} slice exhibits a marginally higher baseline \gls{rtt}. This behavior is an expected artifact of containerized \gls{5g} emulation and the UERANSIM's abstraction of physical-layer \gls{5g} NR (New Radio) over user-space \gls{udp} sockets. Because inter-container forwarding across virtual Ethernet (\texttt{veth}) pairs within the shared host kernel occurs in microseconds, the additional \gls{upf} hops in the \gls{embb} path introduce negligible delay compared to normal inter-core thread scheduling variance across separate \gls{gnb} and \gls{ue} processes.

\begin{figure}[t]
    \centering
    \includegraphics[width=0.75\linewidth]{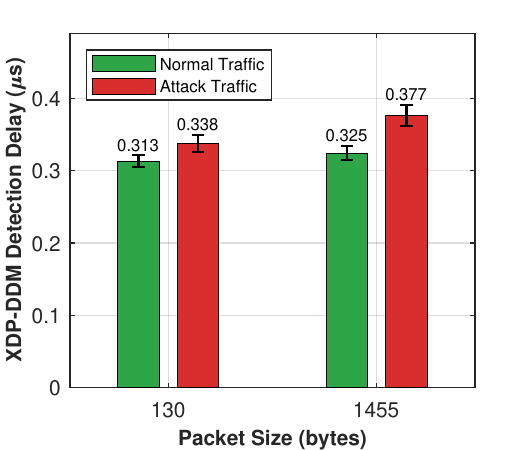}
    \caption{XDP-DDM detection delay ($T_{detect}$) for normal and attack traffic across standard and Maximum Transmission Unit (MTU) packet sizes, averaged over 100 samples ($95\%$ CI).     Algorithm~\ref{alg:xdp_ddm} details the exact computation of this metric.}
    \label{fig:detect}
\end{figure}

When the proposed \gls{ebpf} defense is enabled on the N3 interface, the average \gls{rtt} increases marginally by $1\text{ to }2\text{ ms}$ across both slices, reaching approximately $18.3\text{ ms}$ for the \gls{embb} slice and $20.5\text{ ms}$ for the \gls{urllc} slice. Given that the in-kernel \gls{xdp-ddm} inspection requires less than $0.4\ \mu\text{s}$ per packet (as established in Figure~\ref{fig:detect}), this $1\text{--}2\text{ ms}$ difference reflects normal virtualized network variance across test runs, minor user-space context switching, and the evaluation \gls{icmp} telemetry logging (Algorithm~\ref{alg:xdp_ddm}), rather than data-plane queuing delay. Crucially, the tight $95\%$ confidence intervals confirm that the \gls{xdp} hook introduces no latency spikes or jitter, preserving deterministic performance for latency-sensitive slices.

\begin{figure}[t]
    \centering
    \includegraphics[width=\linewidth]{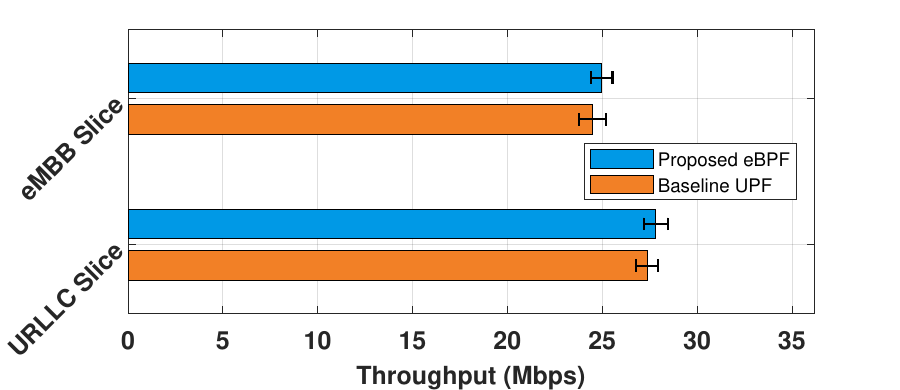}
    \caption{Impact of the proposed eBPF defense on legitimate TCP throughput across eMBB and URLLC slices over 100 samples, where overlapping 95\% confidence intervals demonstrate statistically insignificant overhead compared to the baseline UPF.}
    \label{fig:throughput_overhead}
\end{figure}

\subsubsection{Impact on User Throughput}
Figure~\ref{fig:throughput_overhead} demonstrates the end-to-end \gls{tcp} throughput for legitimate \gls{ue} traffic across the \gls{embb} and \gls{urllc} network slices, comparing the baseline \gls{upf} against the proposed \gls{ebpf} defense. Consistent with the delay evaluation, each bar represents the average throughput computed across 100 clean samples, accompanied by its $95\%$ confidence interval.

Across both network slices, the integration of the proposed \gls{ebpf} program preserves baseline bandwidth performance. Specifically, in the \gls{embb} slice, both the baseline \gls{upf} and the proposed \gls{ebpf} configuration achieve an average throughput of approximately $24.5\text{ to }25\text{ Mbps}$. Similarly, in the \gls{urllc} slice, the throughput remains stable at approximately $27.5\text{ to }28\text{ Mbps}$ under both scenarios. Notably, in contrast to the idle \gls{icmp} \gls{rtt} measurements, the \gls{urllc} slice achieves higher throughput than the \gls{embb} slice. Under continuous, high-volume \gls{tcp} transmissions, the data plane is actively saturated, meaning the cumulative \gls{gtp}-U encapsulation and virtual interface forwarding overhead across the three \gls{upf} hops in the \gls{embb} topology slightly reduces maximum throughput compared to the single-\gls{upf} \gls{urllc} path.

Crucially, when comparing the baseline \gls{upf} to the proposed \gls{ebpf} defense within each respective slice, the $95\%$ confidence intervals overlap substantially. This statistical overlap confirms that the minor difference in average throughput is not statistically significant and represents natural testbed variance rather than \gls{ebpf} processing overhead. Because the \gls{xdp-ddm} performs lightweight, zero-copy header inspection directly within the kernel before packets reach the \gls{upf} forwarding logic, it introduces no noticeable bottlenecks to active \gls{tcp} flows.

In summary, evaluating both end-to-end delay and sustained bandwidth confirms that the proposed \gls{ebpf} security mechanism operates transparently in the data plane, safeguarding the \gls{upf} without degrading the performance of benign slice traffic.

\begin{tcolorbox}[
    colback=blue!7,   
    colframe=blue!50!black,  
    boxrule=0.5pt,           
    arc=3pt,                 
    left=2pt, right=2pt, top=2pt, bottom=2pt 
]
\textbf{Key Takeaway:} The proposed closed-loop \gls{ebpf} defense introduces little to no performance impact on legitimate user traffic. By executing packet inspection directly at the network driver level, the \gls{xdp-ddm} preserves baseline end-to-end latency and throughput across heterogeneous network slices without compromising \gls{qos}.
\end{tcolorbox}

\subsection{Delay of the Closed-Loop Defense} \label{sec_result_latency}

Any security mechanism introduces a performance cost, which in our solution is measured by processing delay. To evaluate the processing efficiency of the proposed \gls{ebpf}/\gls{xdp} defense, we first analyze the delay of each operational phase: detection ($T_{detect}$), reporting ($T_{report}$), and mitigation ($T_{mitigate}$). We assess the detection delay using two packet sizes, a small 130-byte packet and a large 1455-byte packet representing the maximum transmission unit payload.

\textbf{\textit{XDP-DDM Detection Delay:}}
As illustrated in Figure~\ref{fig:detect}, the \gls{xdp-ddm} detection delay is remarkably low, consistently remaining under $0.4\ \mu\text{s}$ across all test scenarios. For 130-byte packets, the detection time for normal traffic averages $0.313\ \mu\text{s}$, while attack traffic requires slightly more time at $0.338\ \mu\text{s}$. A similar marginal gap exists for 1455-byte packets: $0.325\ \mu\text{s}$ and $0.377\ \mu\text{s}$ for normal and attack traffic, respectively.

This slight increase in delay for attack traffic is a direct result of the algorithm's execution path. For benign traffic, the \gls{xdp} program terminates early upon reading the protocol headers. On the other hand, when nested \gls{gtp}-U attack traffic is identified, the algorithm executes additional operations, parsing the inner \gls{udp} header, matching the destination port signature, and extracting the inner source \gls{ip} address prior to performing the BPF LRU map lookup and packet drop.

\begin{figure}[t]
    \centering
    \includegraphics[width=0.75\linewidth]{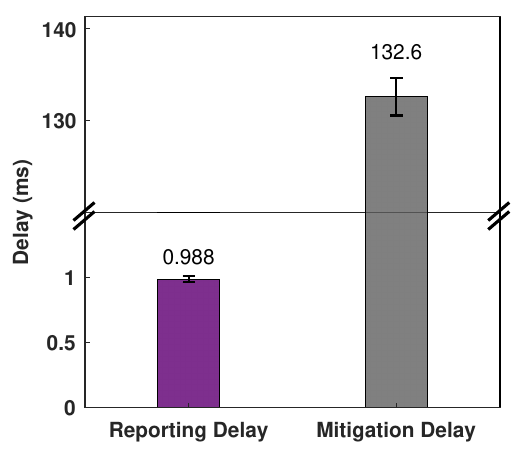}
    \caption{UP-TA telemetry reporting delay ($T_{report}$) shown in purple color and CP-MO mitigation orchestration delay ($T_{mitigate}$) shown in grey color, averaged over 100 samples ($95\%$ CI). Algorithm~\ref{alg:up_ta} and Algorithm~\ref{alg:cp_mo} detail the exact computation of these metrics.}
    \label{fig:contolloop}
\end{figure}

Crucially, while a tenfold increase in total packet size (130 bytes to 1455 bytes) occurs, the detection delay increases by merely $12$ to $40$ nanoseconds. This proves that the \gls{ebpf}/\gls{xdp} program correctly parses headers without buffering or copying the entire payload into memory, maintaining near-constant execution speeds regardless of payload volume. Overall, the detection delay results are highly comparable, with values differing by just under $100$ nanoseconds.

\textbf{\textit{Reporting and Mitigation Delay:}}
Figure~\ref{fig:contolloop} presents the reporting and \gls{cp-mo} mitigation latencies using a broken-axis scale, highlighting the disparity in reporting and core orchestration latencies.

The \gls{up-ta} reporting delay ($T_{report}$) averages just $0.988$ ms. This $1$ ms overhead represents the time required to transfer the monotonic event through the kernel \texttt{trace\_pipe} to user-space (\gls{up-ta}) and parse the resulting JSON alert at the \gls{cp-mo}. This demonstrates that the telemetry dispatch is incredibly lightweight and will not bottleneck the \gls{upf} node's CPU, even during high-frequency alert generation.

In contrast, the \gls{cp-mo} mitigation delay ($T_{mitigate}$) is the most time-consuming phase of the closed-loop defense, averaging $132.6$ ms. This represents the natural networking and processing overhead of executing complex HTTP REST \glspl{api} across the \gls{5g} core, including invoking the \gls{smf} for session release, the \gls{amf} for \gls{ue} deregistration, and the \gls{udr} for the attacker's profile purging.

In summary, while the total closed-loop delay is primarily dominated by the $132.6$ ms mitigation phase, this delay does not expose the victim \gls{ue} to the nested \gls{gtp} attack. Because the \gls{xdp-ddm} has already isolated and dropped the malicious packets in the \gls{dp} within $0.4\ \mu\text{s}$, the victim is instantly protected.  Furthermore, because the \gls{cp-mo} executes the REST API teardown in a separate background thread, it never blocks the reception of new \gls{udp} alerts from the \gls{up-ta}. The system is perfectly asynchronous; the \gls{xdp-ddm} absorbs the nested \gls{gtp} attack instantly, while the \gls{cp-mo} permanently revokes the attacker's network access in the background.

\begin{tcolorbox}[
    colback=blue!7,   
    colframe=blue!50!black,  
    boxrule=0.5pt,           
    arc=3pt,                 
    left=2pt, right=2pt, top=2pt, bottom=2pt 
]
\textbf{\textit{Key Takeaway:}} The proposed defense achieves sub-microsecond ($ < 0.4\ \mu \text{s}$) threat isolation in the data plane, instantly shielding victim \glspl{ue} from nested \gls{gtp} attacks. By asynchronously offloading the control-plane teardown, the system permanently revokes attacker access without introducing blocking overhead or bottlenecks to the \gls{upf} data path.
\end{tcolorbox}

\section{Limitations} \label{sec:limitation}
While the experimental results demonstrate the sub-microsecond efficacy and resilience of the proposed closed-loop \gls{ebpf} defense, several implementation and environmental limitations of this work are as follows:

\textbf{\textit{Hypervisor Clock Skew in Reporting Measurements:}} As noted in Section~\ref{sec:Mitigation_eff}, the measurement of the reporting delay ($T_{report}$) excludes the intermediate virtual switch transit delay. Because the \gls{up-ta} and \gls{cp-mo} execute across isolated virtualized boundaries, inherent hypervisor clock skew introduces sub-millisecond drift between independent system clocks. Consequently, one-way network transit delay across the virtual switch could not be reliably isolated without hardware timestamping and was excluded to preserve measurement accuracy.

\textbf{\textit{Testbed Virtualization and Scalability Bounds:}} Our experimental testbed deploys the \gls{5g} and \gls{ran} components across VirtualBox \glspl{vm} interconnected via paravirtualized (\texttt{virtio-net}) network interfaces. Because UERANSIM emulates the \gls{5g} NR protocol stack in user space over standard \gls{udp} sockets and packet forwarding traverses the host hypervisor's virtual switch, baseline user throughput is bounded to approximately $25-35\text{ Mbps}$, and \texttt{virtio-net} \gls{xdp} execution remains subject to hypervisor context-switching overhead. Consequently, our evaluation focuses on functional efficacy and cross-slice protection under a single-attacker 10,000 \gls{pps} flood rather than saturating \gls{xdp-ddm}'s upper line-rate ceiling. Extensive bare-metal stress testing, including \gls{upf} and \gls{cp} CPU utilization profiling, coordinated multi-\gls{ue} flooding, and comparative benchmarking against alternative filtering approaches, remains an important future direction.

\textbf{\textit{Encrypted N3 Interfaces (IPsec):}} Because \gls{xdp} programs execute at the network driver level prior to the Linux kernel's memory allocation and IPsec decryption stack, the \gls{xdp-ddm} requires visibility into cleartext \gls{gtp}-U headers. Our defense assumes that N3 IPsec tunnels, if deployed, are terminated at an external security gateway prior to reaching the \gls{upf} N3 ingress interface, or decrypted in hardware via SmartNIC inline IPsec offload.

\textbf{\textit{Inner \gls{ip} Spoofing and Outer \gls{teid} Correlation:}} While the \gls{xdp-ddm} extracts and reports both the outer \gls{gtp}-U \gls{teid} and the inner \gls{ip}v4 source address in its telemetry payload, the current defense indexes the 10-second \gls{ebpf} debounce map and performs \gls{cp-mo} session lookup and revocation using only the inner source \gls{ip} address. This design assumes that ingress source-\gls{ip} validation is enforced prior to the \gls{upf} N3 interface (e.g., at the \gls{gnb}). If an adversary can transmit nested \gls{gtp}-U packets with spoofed inner source \glspl{ip}, three secondary risks arise: \textcircled{1} randomized inner \glspl{ip} can bypass the per-\gls{ip} debounce map and flood the telemetry channel, \textcircled{2} unrecognized spoofed \glspl{ip} can evade \gls{cp-mo} session lookup, and \textcircled{3} spoofing another active subscriber's \gls{ip} could trigger an unintended session revocation of an innocent victim \gls{ue}. Because the outer \gls{gtp}-U \gls{teid} is assigned by the trusted serving \gls{gnb} and cannot be forged by the \gls{ue} (unless the attacker controls the \gls{gnb}), basing both the \gls{ebpf} debounce map and the \gls{cp-mo} subscriber resolution strictly on the outer \gls{gtp}-U \gls{teid} resolves all three risks without altering the closed-loop architecture.

\textbf{\textit{Detection Signature and Parsing Scope:}} Our proposed \gls{xdp-ddm} identifies nested \gls{gtp}-U packets by matching inner \gls{udp} traffic destined for the \gls{gtp}-U port 2152. In case a benign \gls{ue} communicates with an external server on port 2152, this check could trigger a false positive. In commercial deployments, this can be eliminated by verifying that the inner destination \gls{ip} address is targeting the operator's core/internal network rather than the \gls{dn}, and by validating the inner \gls{gtp}-U header fields. For the scope of our work, we assume benign \glspl{ue} do not generate non-\gls{gtp} \gls{udp}  traffic destined for port 2152. Additionally, the current solution targets untagged, unfragmented \gls{ip}v4 traffic, while \gls{ip}v6, 802.1Q VLAN tags, and \gls{ip} fragmentation can be accommodated via standard \gls{xdp} parser extensions.

\section{Ethical Considerations} \label{sec:ethical}
All experimental evaluations and the data plane attack in this study were conducted in a localized, privately hosted 5G testbed. To guarantee absolute containment and ensure that no malicious traffic could escape the controlled environment, the Virtual Machines (VMs) hosting the 5G testbed were entirely isolated from the public internet. Furthermore, this work does not introduce any novel attack vectors or zero-day exploits. The implemented attack, specifically the Nested GTP (GTP-in-GTP) attack, relies on previously published and academically acknowledged vulnerabilities. These known exploits were reproduced in our isolated environment solely to empirically validate the defensive efficacy of the proposed framework.

\section{Conclusion} \label{sec:conclusion}
This paper presents a closed-loop \gls{ebpf}/\gls{xdp} security defense designed to mitigate recently uncovered nested \gls{gtp}-U protocol tunneling attacks in \gls{5g} and future \gls{6g} core networks. Empirical evaluations demonstrated that the system neutralizes cross-slice attacks in under $0.4\ \mu\text{s}$, fully preserving the performance of the victim slice. The findings of this study carry several practical implications for the design and operation of \gls{5g}-Advanced and future \gls{6g} data planes. First, our results indicate that integrating lightweight, debounced \gls{ebpf}/\gls{xdp} filters directly at the \gls{upf} ingress provides strong active protection against cross-slice \gls{gtp}-U attacks, allowing operators to safely share underlying \gls{upf} resources across heterogeneous slices. Second, because \gls{xdp-ddm} processes frames within the network driver, the defense operates independently of the specific \gls{upf} software implementation. As standardization bodies transition towards \gls{6g} while retaining \gls{5g} \gls{sba} and \gls{gtp}-U foundations, this plug-and-play defense offers network operators a practical path to harden existing and future Linux-based \glspl{upf} without modifying \gls{vnf} source code.

\section*{Acknowledgment}
This work was supported by the Natural Sciences and Engineering Research Council of Canada (NSERC). This research was enabled in part by support provided by Research Computing Services (https://carleton.ca/rcs) at Carleton University.

\section*{Generative AI Usage Considerations}
We, the authors, are fully responsible for the originality and correctness of all content in this paper, including the contributions, proposed defense, evaluation methodology, experimental results, and conclusions. Generative AI was used for editorial purposes in this manuscript, and all outputs were inspected by the authors to ensure accuracy and originality.

\bibliographystyle{ieeetr}
\bibliography{library}

\end{document}